# Over-8-dB squeezed light generation by a broadband waveguide optical parametric amplifier toward fault-tolerant ultra-fast quantum computers

Takahiro Kashiwazaki,[1),a)] Taichi Yamashima,[2)] Koji Enbutsu,[1)] Takushi Kazama,[1)] Asuka Inoue,[1)] Kosuke Fukui,[2)] Mamoru Endo, [2),3)] Takeshi Umeki,[1)] and Akira Furusawa[2),3),b)]

[1)] *NTT Device Technology Labs, NTT Corporation, 3-1, Morinosato Wakamiya, Atsugi, Kanagawa, 243-0198, Japan*

[2)] *Department of Applied Physics, School of Engineering, The University of Tokyo, 7-3-1 Hongo, Bunkyo-ku, Tokyo, 113-8656, Japan*

[3)] *Optical Quantum Computing Research Team, RIKEN Center for Quantum Computing, 2-1 Hirosawa, Wako, Saitama, 351-0198, Japan*

[a)] Author to whom correspondence should be addressed: takahiro.kashiwazaki.dy@hco.ntt.co.jp
[b)] Electronic mail: akiraf@ap.t.u-tokyo.ac.jp

**ABSTRACT**

We achieved continuous-wave 8.3-dB squeezed light generation using a terahertz-order-broadband waveguide optical parametric amplifier (OPA) by improving a measurement setup from our previous work [T. Kashiwazaki, et al., Appl. Phys. Lett. **119**, 251104 (2021)], where a low-loss periodically poled lithium niobate (PPLN) waveguide had shown 6.3-dB squeezing at a 6-THz frequency. First, to improve efficiency of the squeezed light detection, we reduced effective optical loss to about 12% by removing extra optics and changing the detection method into a low-loss balanced homodyne measurement. Second, to minimize phase-locking fluctuation, we constructed a frequency-optimized phase-locking system by comprehending its frequency responses. Lastly, we found optimal experimental parameters of a measurement frequency and a pump power from their dependences for the squeezing levels. The measurement frequency was decided as 11 MHz to maximize a clearance between shot and circuit noises. Furthermore, pump power was optimized as 660 mW to get higher squeezing level while suppressing anti-squeezed-noise contamination due to an imperfection of phase locking. To our knowledge, this is the first achievement of over-8-dB squeezing by waveguide OPAs without any loss-correction and circuit-noise correction. Moreover, it is shown that the squeezing level soon after our PPLN waveguide is estimated at over 10 dB, which is thought to be mainly restricted by the waveguide loss. This broadband highly-squeezed light opens the possibility to realize fault-tolerant ultra-fast optical quantum computers.

Fault-tolerant universal quantum computers using continuous-variables (CVs) of traveling wave of light are expected to be the fastest computers in the world, executing not only quantum algorithms but also classical ones with terahertz-order clock frequency [1, 2]. By applying measurement-based quantum computing (MBQC) [3] with a time-domain-multiplexed optical quantum entangled state [4, 5, 6], we can perform large-scale quantum computation without expanding or integrating various apparatus [7]. Furthermore, MBQC with flying qubits does not need to care about a coherence time of the qubits. This is because qubits are all used before their decoherence, since local measurements in MBQC perform quantum calculation and collapse the wave function.

To achieve fault-tolerant CV optical MBQC with high-clock-frequency, broadband high-level squeezed light is required as the most important quantum resources. Broad bandwidth helps us to define shorter quantum wave packets in a narrow time range. It leads to large-scale and high-clock-frequency quantum computers using a time-domain multiplexing technique. Furthermore, broadband squeezers have been shown to be capable of semi-deterministic generation for non-Gaussian-states [11, 12], which are essential ancillary states to achieve universal and fault-tolerant MBQC with high-clock-frequency. As for the squeezing level, high-level squeezing contributes to high-fidelity quantum operation and suppressing quantum errors. For example, fault-tolerant CV MBQC with Gottesman-Kitaev-Preskill (GKP) coding [14, 15] requires more than about 8-dB squeezing [16, 17].

To achieve both high-level and broadband squeezing at the same time, we have developed $\chi^{(2)}$-based waveguide optical parametric amplifiers (OPAs) [18, 19]. Thanks to their light confinement in a small core, waveguide OPAs can show relatively higher nonlinear effects without using optical cavities. This allows single-pass squeezed light generation resulting in broad bandwidth [20, 21, 22, 23, 24]. We reported in 2020 that a directly-bonded periodically poled lithium niobate (PPLN) waveguide achieved continuous-wave (CW) 6.3-dB squeezing at 20-MHz sideband frequency thanks to its spatial-single-mode structure and high-durability against high-power pump light [18]. This was the first realization of CW squeezing with a single-pass OPA at a level exceeding 4.5 dB without any loss correction, which is required for the generation of a two-dimensional cluster state. Furthermore, this paper showed that the squeezing-level deterioration due to gain-induced phenomena and contamination from anti-squeezed noise of higher-ordered spatial modes can be overcome by single-spatial-mode waveguide structure. The paper also showed that in order to increase squeezing levels we should focus on reducing optical losses. In 2021 an improved low-optical-loss PPLN waveguide showed 6.3-dB squeezing at over-6-THz bandwidth by using an all-optical quadrature measurement in a fiber-closed setup [19]. However, the measured squeezing level from an OPA has not yet reached the level required for GKP error correction. Because Ref. 19 focused on THz-order broadband measurement rather than high-level squeezing, the detection efficiency was not so high. The effective optical loss for the squeezed light was up to 21%.

In this letter we report achievement of 8.3-dB quadrature squeezing at an 11-MHz sideband frequency by removing extra optics and changing the detection method into a low-loss balanced homodyne measurement from our previous paper [19]. We use a low-loss THz-order-bandwidth OPA consisting of a PPLN waveguide, which is fabricated by directly-bonding [25] and mechanical-sculpturing processes [26]. In addition, to minimize the effect of phase locking fluctuation, we construct a frequency-optimized phase-locking system from the frequency analysis of the system response. Furthermore, we decide an optimal measurement frequency and pump power by measuring frequency- and pump-power-dependences of the squeezing levels. The measurement frequency was decided as 11 MHz to maximize a clearance between shot and circuit noises. Furthermore, pump power was optimized as 660 mW to get higher squeezing level while suppressing anti-squeezed-noise contamination due to an imperfection of phase locking. To our knowledge, this is the first achievement of over-8-dB squeezing by THz-order broadband OPAs without any loss correction and circuit-noise correction. Besides, by subtracting detection loss of the homodyne measurement, we estimate the squeezing level soon after our PPLN waveguide as over 10 dB. This value is good agreement with the expected values mentioned in our previous work [19].

Figure 1(a) shows a conceptual image of our previous measurement setup showing 6.3-dB 6-THz squeezing in Ref. 19. A fiber-pigtailed OPA module contains a 45-mm-long PPLN waveguide and has two input polarization-maintaining optical fibers for pump light and probe light. The PPLN waveguide is fabricated by a mechanical polishing process [19, 26], which results in optical loss lower than that of dry-etched waveguides. The PPLN waveguide shows a second-harmonic-generation (SHG) coefficient of 820 %/W and a waveguide loss of 7% [19]. The detection was done by an all-optical quadrature measurement technique to measure the squeezing level of over THz-order bandwidth. In this measurement, the effective optical loss was 21%, which consists of about 4% for waveguide loss, 8% for output side of the OPA module, and 10% for an input side of the amplification OPA. The last one is known as a noise figure of OPAs [27, 28]. To detect higher squeezing level, we improve the detection efficiency by reducing extra optics in the output side of the OPA module and changing the detection method into the low-loss balanced homodyne measurement shown in Fig. 1(b).

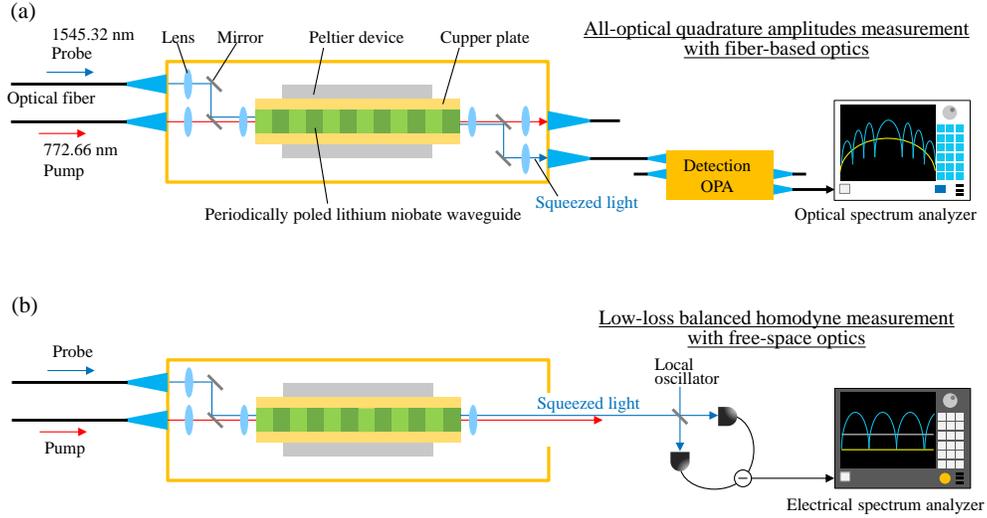

Fig. 1 (a) Schematic view of our previous experiment [19]. Squeezed light from a fiber-pigtailed optical parametric amplifier (OPA) module is detected by an all-optical quadrature amplitudes measurement. The OPAs contain low-loss PPLN waveguides. The squeezed noise level was measured by an optical spectrum analyzer. (b) A modified OPA module for measuring higher squeezing level. Squeezed light is output from the OPA into free space and detected by a low-loss balanced homodyne measurement. The noise level is measured by an electrical spectrum analyzer.

The detailed experimental setup for the squeezing level measurement is shown in Fig. 2(a). First, we explain optical paths. Since squeezed light does not have certain classical optical amplitude and phase, we have to use probe light for phase locking between squeezed light and a local oscillator (LO). Continuous-wave (CW) laser light with a wavelength of 1545.32 nm is emitted from a fiber-laser (NKT Photonics, BASIK module). The light is amplified by an erbium-doped optical fiber amplifier (EDFA) (Keyopsys, CEFA-C-PB-HP). The amplified light is split into mainly three optical lines, pump light for squeezed light generation, probe light, and the LO. To generate pump light for squeezed light generation, we use an optical frequency doubler (NTT Electronics, WH-0772-000-F-B-C) consisting of a PPLN waveguide. The frequency-doubled beam is injected into our PPLN-based OPA. Inside the OPA, squeezed light is generated according to optical parametric down conversion process and propagates with probe light. After the OPA, the squeezed light travels in free space. The probe light has been frequency shifted by two serially-connected acousto-optic-modulators (AOMs) (Chongqing Smart Science &Technology Development, SGTF-40-1550-1P). The shift frequency is 1 MHz, which is determined by Bode spectra obtained from system identification of the phase-locking circuits. The details of this analysis are shown later in this paper. After the OPA, the squeezed light is reflected by a dichroic mirror (DM1), while residual pump light goes thorough DM1. One percent power of squeezed and probe light is separated by a 99/1 mirror and detected by an avalanche photodiode (PD2) (Thorlabs, APD-430C). This light intensity is used for phase locking between probe light and squeezed light. The other 99% of the light from the 99/1 mirror is detected by a balanced homodyne detector (BHD), which has two InGaAs photodiodes (Laser Components, IGHQEX0100-1550-10-1.0-SPAR-TH-40). The quantum efficiencies of these photodiodes are both about 98%. A fringe visibility between probe light and the LO is about 98.5%, which results in optical loss of about 3%. Optical loss of the squeezed-light path is measured about 3% including 1%-power dividing at the 99/1 mirror for phase-locking. Therefore, total optical loss for squeezed light output from the OPA is estimated at about 8%. The signal from the detector is analyzed by an electrical spectrum analyzer (Keysight, N9010B) (ESA). In the paths for the LO and probe light, fiber-stretchers (FSs) and electro-optic modulators (EOMs) are inserted for phase locking.

Next, we explain electrical paths for phase locking. The signal from PD2 is frequency-down-converted by 2 MHz with the use of an electrical mixer and an electrical low-pass filter. The down-converted signal is injected into a Proportional-Integral-Derivative (PID) controller as an error signal for phase locking between the probe and the squeezed light. The control signal is connected to the EOM and to another PID controller. The second PID controller generates a slow control signal for the FS, which can tune optical length within a relatively long range. This configuration enables us to lock the relative phase between probe and squeezed light. With respect to the phase locking between the probe and the LO, we use a signal from the BHD and two other PID controllers. The signal is frequency-down-converted by 1 MHz. Here, these down converting frequencies are decided by Bode spectra of the phase-locking system. The frequency is preferred to be higher, unless the phase locking system oscillates. To measure the system responses, such as a loop gain and a phase delay, we use a network analyzer (Keysight, E5061B). By replacing a dashed box of A (or B) into that of C in Fig. 2(a), we insert the network analyzer in the place between the electrical mixer and the EOM. During system response measurements, the second PID and FS of the loop under test with relatively slow control are used to maintain the phase-locked state. This measurement enables us to know the system response of higher frequency than phase-locked frequency by the second PID [29]. Figures 2(b) and 2(c) show the Bode spectra of the phase-locking system for squeezed light and the LO. The phase delays of the systems for squeezed light and the LO reach to −180 degrees when the frequencies are about 4 MHz and 2 MHz, respectively. Furthermore, the gains are almost stable under the 1 MHz frequency. From these results, we decide the frequency shift of 1 MHz at two AOMs. The spikes at 2 MHz in Fig. 2 (b) and at 1 MHz in Fig. 1(c) are due to the frequency shift at AOMs of 1 MHz for phase locking by the second PID. Note that since the probe light induces a differential-frequency-generation process in the OPA, the beat frequency becomes twice the original shift frequency of 1 MHz at AOMs.

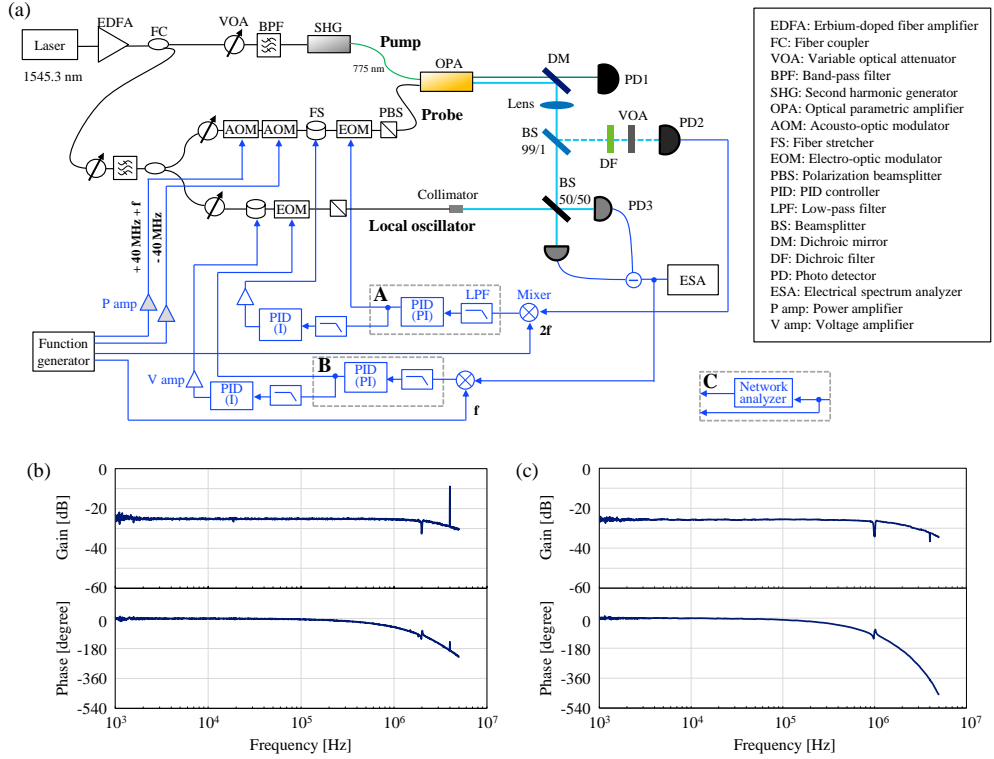

Fig. 2. (a) Schematic diagram of an experimental setup for detection of squeezed light from a PPLN waveguide. For phase locking, we inject probe light into the PPLN waveguide. When we measured system responses, the dashed box A (or B) was exchanged for the dashed box C containing a network analyzer. (b) Bode diagrams of the phase-locking system between squeezed light and probe light. The upper graph is a system gain, and the lower one is a phase delay. (c) Bode diagrams of the phase-locking system between the probe light and the LO.

Figure 3 shows squeezing levels detected by the spectrum analyzer in zero span mode. The measurement center frequency is 11 MHz, a resolution bandwidth is 1 MHz, and a video bandwidth is 1 kHz. The pump power in the PPLN waveguide is 660 mW. These parameters of the measurement frequency and the pump power are optimized from the later analysis for frequency and pump-power dependence of squeezing levels. The power of the LO soon before the BHD is 15 mW. The phase-locked squeezed noise is 8.3±0.1 dB lower than the shot noise level without any loss correction and circuit-noise correction. This squeezing level reaches the error-correctable threshold with GKP coding [16, 17].

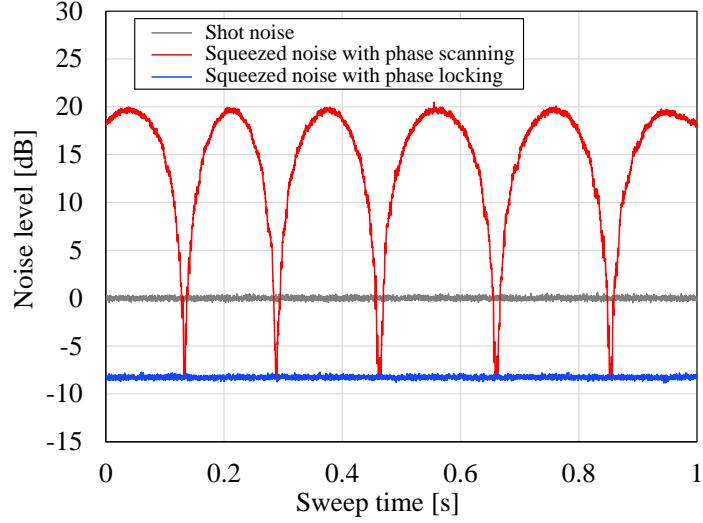

Fig. 3. Noise levels detected by a spectrum analyzer in zero span mode. The measurement center frequency is 11 MHz with a resolution bandwidth of 1 MHz and a video bandwidth of 1 kHz. There are two types of squeezed noise are shown. One with phase scanning of the local oscillator and the other with phase locking.

In the following paragraphs, we explain how to decide the measurement center frequency and the pump power by frequency and pump-power dependence of squeezing levels. Figure 4(a) shows frequency dependence of squeezed noise detected by the spectrum analyzer with a resolution bandwidth of 1 MHz. Circuit noise of the detector, shot noise, and floor noise of the spectrum analyzer are also shown. The clearance between shot noise and circuit noise reaches the large value of about 25 dB around 11 MHz. This circuit noise is equivalent to the optical loss of about 0.3% for squeezed light. At higher frequency, the squeezing level is reduced according to the circuit noise increment. Therefore we decided the measurement frequency as 11 MHz.

As for the pump-power dependence, the squeezing and anti-squeezing levels are shown in Fig. 4(b). Theoretical curves are drawn by the following equation;[30, 31]

$$R'_\pm(\tilde{\theta}) \approx R_\pm \cos^2 \tilde{\theta} + R_\mp \sin^2 \tilde{\theta}. \quad (1)$$

Here $\tilde{\theta}$ is a standard deviation of relative phase angle difference between the LO and anti-squeezed or squeezed quadratures with assuming small angle. $R_+$ and $R_-$ are actual anti-squeezed and squeezed noise levels, which are described in following equation. [19, 20, 21]

$$R_\pm = (1-\eta) + \eta \cdot \exp(\pm 2\sqrt{\alpha P}). \quad (2)$$

Here $\eta$ is effective optical transmittance of squeezed light, $\alpha$ is SHG efficiency, and $P$ is pump power. To draw the theoretical curves, we use an $\alpha$ of 820 %/W as in Ref. 19. In addition, we assume $\eta$ as 12% by summing of detection optical loss of 8%, equivalent loss from circuit noise of 0.3%, and effective loss of inside the waveguide as 4% [19]. The theoretical curve without phase angle fluctuation ($\tilde{\theta} = 0$) is shown as a dashed line. This ideal line shows that the squeezing level improves according to pump power increment. However, an actual plot shows the degradation of squeezing levels in the high-pump-power region. The squeezing level is headed with a pump power of around 660 mW, which is thought to be an optimal pump power of this experimental setup. We think this degradation in the high-power region is mainly due to the locked-phase fluctuations, although there would be various reasons of this degradation, such as pump-induced phenomena. As a reason for this, the measured squeezing and anti-squeezing levels are well fitted to the theoretical curve drawn assuming $\tilde{\theta}$ is 0.8 degrees, which is shown by the solid line in Fig. 4 (b). Therefore, for higher squeezed light generation, we have to minimize the phase fluctuation as well as to reduce the optical loss as future works. In addition, the good fit of the theoretical curve means that the parameters of the optical loss and the SHG coefficient are reasonable values for the actual conditions. By subtracting the 8% optical loss in the homodyne detection, more than 10 dB of squeezing is expected soon after the PPLN waveguide. This result is consisting with the estimated value in Ref. 19. We thought that this squeezing level is restricted by the optical loss of the waveguide.

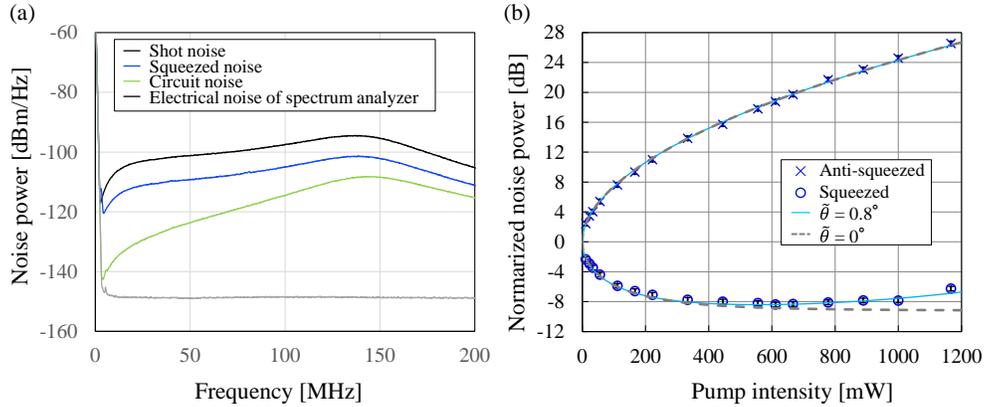

Fig. 4. (a) Frequency dependence of the squeezed noise detected by a spectrum analyzer. A resolution bandwidth is 1 MHz, and a video bandwidth is 1 kHz. Shot noise, circuit noise, and electrical noise of the electrical spectrum analyzer are also shown. (b) Pump-power dependence of squeezing and anti-squeezing levels. We draw theoretical curves assuming phase fluctuation angles as 0.8 degrees for the solid line and as 0 degrees for the dashed line.

In conclusion, we achieved 8.3-dB quadrature squeezing at 11-MHz sideband frequency by using a low-loss THz-order-bandwidth OPA consisting of a PPLN waveguide. The PPLN waveguide was fabricated by directly-bonding and mechanical-sculpturing processes to achieve low optical loss. To detect higher squeezing levels, we reduced effective optical loss to about 12% by removing extra optics and changing the detection method into a low-loss balanced homodyne measurement from our previous paper [19]. In addition, to minimize the effect of phase-locking fluctuation, we constructed a frequency-optimized phase-locking system with the shift frequency of the probe light of 1 MHz, which was decided by the Bode analysis for the system. Furthermore, we found an optimal measurement frequency and pump power were 11 MHz and 660 mW by frequency- and pump-power-dependent of the squeezing and anti-squeezing levels. To our knowledge, this is the first achievement of over-8-dB squeezing by THz-order broadband OPAs without any loss correction and circuit-noise correction. This squeezing level reaches the error correctable threshold with GKP coding. Besides, the pump-power-dependent squeezing and anti-squeezing levels were well fitted to the theoretical curve assuming optical loss for squeezed light as 12% and SHG efficiency as 820 %/W, which are good agreement with the measurement results of actual optical components and our previous paper [19]. Furthermore, the phase locking fluctuation was estimated as at most 0.8 degrees. These results indicated that the squeezing level soon after our PPLN waveguide is estimated at over 10 dB. This broadband highly-squeezed light opens the possibility to realize fault-tolerant ultra-fast optical quantum computers with GKP coding.

## ACKNOWLEDGEMENT

The authors acknowledge supports from the UTokyo Foundation and donations from Nichia Corporation of Japan. T.Y. acknowledges support from the Advanced Leading Graduate Course for Photon Science (ALPS). M.E. acknowledges support from the Research Foundation for Opto-Science and Technology. This work was partly supported by the Japan Science and Technology Agency (JPMJMS2064, JPMJPR2254) and the Japan Society for the Promotion of Science KAKENHI (18H05207, 20K15187).## DATA AVAILABILITY

The data that support the findings of this study are available from the corresponding author upon reasonable request.

## DISCLOSURES

The authors declare no conflicts of interest.

## REFERENCES

1. W. Asavanant and A. Furusawa: Optical Quantum Computers A Route to Practical Continuous Variable Quantum Information Processing. AIP Publishing, Melville, New York (2022)

2. S. Takeda and A. Furusawa, "Toward large-scale fault-tolerant universal photonic quantum computing," APL Photonics **4**, 060902 (2019).

3. R. Raussendorf and H. J. Briegel, "A one-way quantum computer," Phys. Rev. Lett. **86**, 5188 (2001).

4. S. Yokoyama, R. Ukai, S. C. Armstrong, C. Sornphiphatphong, T. Kaji, S. Suzuki, J. I. Yoshikawa, H. Yonezawa, N. C. Menicucci, and A. Furusawa, "Ultra-large-scale continuous-variable cluster states multiplexed in the time domain," Nat. Photonics **7**, 982 (2013).

5. W. Asavanant, Y. Shiozawa, S. Yokoyama, B. Charoensombutamon, H. Emura, R. N. Alexander, S. Takeda, J. Yoshikawa, N. C. Menicucci, H. Yonezawa, and A. Furusawa, "Generation of time-domain-multiplexed two-dimensional cluster state," Science **366**, 373 (2019).

6. W. Asavanant, B. Charoensombutamon, S. Yokoyama, T. Ebihara, T. Nakamura, R. N. Alexander, M. Endo, J. Yoshikawa, N. C. Menicucci, H. Yonezawa, and A. Furusawa, "Time-Domain-Multiplexed Measurement-Based Quantum Operations with 25-MHz Clock Frequency," Phys. Rev. Applied **16**, 034005 (2021).

7. K. Fukui and S. Takeda, "Building a large-scale quantum computer with continuous-variable optical technologies," J. Phys. B: At. Mol. Opt. Phys. **55**, 012001 (2022)

8. F. Arute, et al., "Quantum supremacy using a programmable superconducting processor," Nature **574**, 505 (2019).

9. J. M. Pino, et al., "Demonstration of the trapped-ion quantum CCD computer architecture," Nature **592**, 209 (2021).

10. T. F. Watson, et al., "A programmable two-qubit quantum processor in silicon," Nature **555**, 633 (2018).8